\documentclass[article,twocolumn,showpacs,preprintnumbers,amsmath,amssymb]{revtex4}
\usepackage{graphicx}
\usepackage{dcolumn}
\usepackage{bm}
\usepackage{epsfig}

\begin{document}

\title[Formation time of resonances]{Formation and decay of hadronic resonances
in the QGP}

\author{C. Markert$^1$, R. Bellwied$^2$, I. Vitev$^3$}

\vspace*{.5cm}

\affiliation{$^1$ Physics Department, University of Texas at Austin,
Austin, TX 78702, USA}
\email{cmarkert@physics.utexas.edu}

\affiliation{$^2$ Physics Department, Wayne State University,
666 West Hancock, Detroit, MI 48201, USA}

\affiliation{$^3$ Los Alamos National Laboratory, Theoretical Division, Mail Stop
B283, Los Alamos, NM 87545, USA}

\vspace*{1cm}

\begin{abstract}

Hadronic resonances can play a pivotal role in providing
experimental evidence for partial chiral symmetry restoration in the
deconfined quark-gluon phase produced at RHIC. Their lifetimes,
which are comparable to the lifetime of the partonic plasma phase,
make them an invaluable tool to study medium modifications to the
resonant state due to the chiral transition. In this paper we show
that the heavier, but still abundant, light and strange quark
resonances $K^*, \phi$, $\Delta$  and $\Lambda^*$ have large
probability to be produced well within the plasma phase due to their
short formation times. We demonstrate that, under particular
kinematic conditions, these resonances can be formed and will decay
inside the partonic state, but still carry sufficient momentum to
not interact strongly with the hadronic medium after the QCD phase
transition. Thus, $K^*, \phi$, $\Delta$ and $\Lambda^*$ should
exhibit the characteristic property modifications which can be
attributed to chiral symmetry restoration, such as mass shifts,
width broadening or branching ratio modifications.

\end{abstract}


\maketitle


\section{Introduction}

Measurements of experimental observables in hadron-nucleus
collisions at relativistic energies strongly suggest that
final-state hadrons are not produced at the moment of collisions
$\tau_{coll.} \sim R_A/\gamma $,  $\gamma = E_N/m_N$. Instead, they
require a finite formation time, which can be viewed as the time
needed to build up the hadronic wave function, i.e. the time to
establish wavefunction overlap between the constituent partons.
Based on the uncertainty principle and time dilation, the production
of relativistic particles is supposed to follow a simple particle
energy dependence, i.e. low energy particles are produced first,
high energy particles - last. In elementary particle physics this is
known as the 'inside-outside cascade'. Its main features have been
verified in many deep inelastic scattering experiments~\cite{bial}.
The measured dependence of hadron production modification on the
mass of the nuclear target shows how important the interplay of this
simple hadronization time dependence and the space-time evolution of
the parton-nucleus interaction can be. Several model calculations
have argued that 'pre-hadron' formation is a possible mechanism that
can account for the scaling  of hadron multiplicity ratios in cold
nuclear matter~\cite{accardi,cassing}.

In relativistic heavy ion collisions (A+A) the hadron formation time is of
even greater importance because, here, a portion of the hadrons are
likely formed during the partonic lifetime of the quark-gluon plasma
(QGP), and therefore will generate a mixed phase with the QGP. The
interactions of these hadrons with the plasma phase could
potentially be used to elucidate both fundamental physics concepts
and specific physics mechanisms, such  as the existence of chiral
symmetry restoration in high energy nuclear collisions or the nature
of heavy flavor energy loss in the plasma.  Adil and Vitev~\cite{vitev}
have recently shown that early hadronization of heavy quarks
triggers a new type of interaction between the charm and beauty mesons
and the hot partonic medium. Collisional dissociation of heavy
mesons in the QGP emulates large energy loss 
and leads to significant suppression of their experimentally observed
cross sections, as measured through semi-leptonic decay channels by
STAR and PHENIX~\cite{heavy1,heavy2}.

In this paper we build upon the recent developments toward
incorporating the space-time picture of hadronization in heavy ion
phenomenology and evaluate the formation time of light- and
strange-quark heavy-mass hadronic resonances, such as the K$^{*}$,
$\phi$, $\Delta$ and $\Lambda^{*}$, in order to determine the
probability of interaction between early-formed resonances and the
partonic medium. In the second part of the paper we propose, based
on the encouraging model results, a specific triggered jet
correlation measurement for RHIC and the LHC, which could resolve
the question of resonance
modification~\cite{Holt:2004tp,Filip:2001st,Cassing:1997jz}, including baryon
states~\cite{baryons}, through partial chiral symmetry restoration.

\section{Treatment of formation time in high energy collisions}

The simplest formulation of the energy and mass dependence of the
inside-outside cascade~\cite{Bjorken:1976mk} can be written as:
\begin{equation}
 \tau_{\rm form} = \tau_0 \frac{E}{m} \; ,
\label{iocas}
\end{equation}
where $\tau_{\rm form}$ is the formation time, $\tau_0$ is
the proper formation time in the hadron's rest frame, $E$
is the energy of the hadron and $m$ is its mass. Eq.~(\ref{iocas})
has the transparent interpretation of
a Lorentz-dilated interval $\tau_0 \sim R_h \sim 1$~fm/c. In such a
model high energy particles are produced later, and heavy mass
particles are produced earlier.

The space-time picture of hadronization was re-examined for
string fragmentation models, such as the Lund model~\cite{lund},
to point out that different hadron constituents are produced at
different times~\cite{Kopeliovich:1984ur,gyulassy}.  One can then distinguish
between a constituent length, which is the time ($c=1$)
after which the valence quarks of the hadron are formed through
color string breaking, and the formation length,
which is the time when the quarks coalesce to form
a hadron. Between constituent length and formation length the
structure can be envisioned as a pre-hadron. The constituent
length itself vanishes when the hadron is near the kinematic
limit $z \rightarrow 1$ and this inverted formation hierarchy
is known as  an outside-inside cascade. If final-state hadron
(or pre-hadron)
absorption in cold nuclear matter is the origin of the observed
cross section attenuation in the HERMES semi-inclusive DIS
measurements~\cite{falter}, constituent length would appear
to more accurately capture the dynamics of meson and baryon
production in high energy collisions.

In this paper we are primarily interested in the production of
leading strange-quark resonances and their triggered correlations.
For such special kinematics and strong bias towards large
momentum fractions $z = p^+_{\rm hadron}/p^+$  models of cascade
hadronization can only serve as a guidance. In view of this, let us
revisit the formation time evaluation~\cite{vitev}, which is most easily
understood in the context of independent fragmentation~\cite{akk,soffer}.
In momentum  space, in lightcone coordinates, the fragmentation of
a parton of mass $m_q$ into a hadron of mass $m_h$ and a
light secondary parton can be represented as:
\begin{eqnarray}
\left[p^+, \frac{m_q^2}{2p^+}, {\bf 0} \right]
& \rightarrow & \left[zp^+, \frac{ {\bf k}^2 +  m_h^2}{2zp^+},
{\bf k} \right]  \nonumber \\
&& +
 \left[(1-z)p^+, \frac{ {\bf k}^2 }{2(1-z)p^+},
-{\bf k} \right] \;.  \quad
\label{fragtime}
\end{eqnarray}
Here $p^+$ is the large lightcone momentum
of the parent parton and $|{\bf k}| \sim \Lambda_{QCD}
\sim 200$~MeV/c  is the deviation from collinearity.
We can evaluate  the  lightcone  $\Delta y^+ =  \tau_{\rm form} +
l_{\rm form}$   that is conjugate to the non-conserved lightcone
momentum component
$\Delta p^- = (p^-)_{final} - (p^-)_{initial} $ of the evolving system:
\begin{eqnarray}
\Delta y^+ &\simeq& \frac{1}{\Delta p^-}  \;
\nonumber \\
& = & \; \frac{zp^+}{m_h}   \times  2 \left[ m_h +
\frac{ {\bf k}^2}{(1-z)m_h} - \frac{zm_q^2}{m_h}
                              \right]^{-1} \;. \quad
\label{tfrag}
\end{eqnarray}
The formation time then reads:
\begin{eqnarray}
&& \tau_{\rm form} = \frac{ \Delta y^+}{ 1+\beta_q} \;, \quad
\beta_q = \frac{p_q}{E_q} \;.
\label{tform}
\end{eqnarray}
We note that Eq.~(\ref{tform}) exhibits features of both inside-outside
and outside-inside cascades. It can be interpreted, similarly to
Eq.~(\ref{iocas}), as a boost to a proper formation time, which now
depends on the details of the parent parton decay. Our results, however,
also shows that $\tau_{\rm form} \rightarrow 0$ when $z \rightarrow 1$.
This effect is most pronounced for light pions, where
$m_\pi < \Lambda_{QCD}$, and there is absolute scale sensitivity
to the choice of the non-perturbative physics parameter
$|{\bf k}| \sim \Lambda_{QCD}$. For heavy  resonances such
sensitivity and the kinematic region of outside-inside cascade
are strongly reduced.

\section{QGP evolution}

At present there is no reliable dynamical mechanism that can
quantitatively describe the formation of a deconfined state, the
quark-gluon plasma, in ultra-relativistic heavy ion collisions.
Nuclear stopping, which at the partonic level is described by
inelastic quark and gluon interactions in cold nuclear
matter~\cite{Vitev:2007ve}, plays a major role in the liberation of
partonic degrees of freedom. However, for center of mass energies
$\sqrt{s}_{NN} > \sqrt{s}_{SPS} \approx 17$~GeV the collision time
$\tau_{coll.} \simeq 2 R_A / \gamma $ is  generally considered
insufficient for isotropisation and/or equilibration of the QGP,
should such state of matter be formed. For example, in central Pb+Pb
collisions at the LHC  $\tau_{coll.} = 5 \times 10^{-3}$~fm/c
($\rightarrow 0$) and can be used to mark the beginning of
final-state evolution with very little uncertainty. We can gain
insight in the timescale of  medium formation using the approximate
parton-hadron duality~\cite{Vitev:2005he} and the uncertainty
principle.  In high energy hadronic collisions light pions dominate
multiplicities  and $\tau_0 \approx 1 / \langle p_T \rangle_\pi $.
In our evaluation we take $\langle p_T \rangle  = 450$~MeV/c for
central Au+Au collisions at RHIC~\cite{Adler:2003cb} and note that
there is approximately $25\%$ variation of the mean transverse
momentum in going from central to peripheral collisions.
Extrapolations to LHC energies have been made using Monte Carlo
event generator results, fit to the CDF collaboration data from $
\sqrt{s} = 1.8 $~GeV $p+\bar{p}$ collisions~\cite{Sjostrand:2006za}.
Accounting for the above-mentioned variation in going from $N+N$  to
$A+A$ collisions we find $ \langle p_T \rangle  = 850  $~MeV/c for
central Pb+Pb collisions at the LHC. Consequently, $\tau_0^{RHIC} =
0.44$~fm/c and $\tau_0^{LHC} = 0.23$~fm/c.

Gluon degrees of freedom (8 colors, 2 polarizations) dominate soft
parton multiplicities at RHIC and the LHC  and their density can be
related to charged hadron rapidity density in a Bjorken  expansion
scenario as follows~\cite{Vitev:2005he}:
\begin{equation}
\rho(\tau) =  \frac{1}{\tau A_\perp} \frac{dN^g}{dy}
\approx  \frac{1}{ \tau A_\perp} \frac{3}{2}
\left| \frac{d\eta}{dy} \right|  \frac{dN^{ch}}{d\eta} \; .
\label{ydep}
\end{equation}
Assuming local thermal equilibrium one finds:
\begin{equation}
T(\tau) = \  ^3\!\sqrt{ \pi^2 \rho(\tau) / 16 \zeta(3) } \; .
\label{tempdet}
\end{equation}
In Eqs.~(\ref{ydep}) and (\ref{tempdet}) the average over the
distribution of densities, and correspondingly temperatures, in
the plane transverse to the collision axis is implicit.
For $b=3$~fm, $dN^g/dy = 1100$ at RHIC and $dN^g/dy = 2800$ at
the LHC  in Au+Au and Pb+Pb collisions, respectively,
we find $T(\tau_0)_{RHIC} = 435$~MeV and
$T(\tau_0)_{LHC} = 713$~MeV.

The lifetime of the QGP itself depends on the details of the
expansion. Different hydrodynamic scenarios, including initial
conditions and type of expansion, appear to be approximately
consistent with the low $p_T$ data. In general, the faster the
transverse expansion $v_T \neq 0$, the shorter the lifetime of
the plasma will be. Thus, we can reliably give only the upper time
limit of the transition from the QGP phase to a hot hadron gas,
corresponding to the case of longitudinal Bjorken expansion:
\begin{equation}
\tau_{QGP} = \tau_0 (T_0/T_c)^3
\end{equation}
For a critical temperature $T_c \approx 180$~MeV, we find at RHIC
$\tau_{QGP} = 6.2$~fm/c and at the LHC $\tau_{QGP} = 14$~fm/c. We
emphasize again that these are mean values over the reaction
geometry and the hottest densest central region of the collision
will have larger QGP lifetimes. On the other hand, transverse expansion
will reduce the QGP lifetime, bringing the RHIC estimate in accord with
the measurement of hadronic resonance suppression in conjunction
with HBT measurements \cite{markert1}.

\begin{figure}
\includegraphics[width=2.9in]{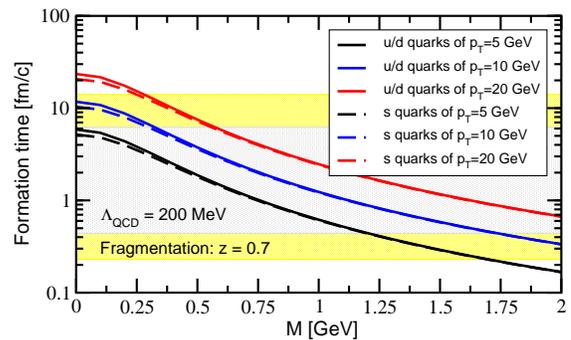}
\label{formM} \vspace*{.2cm} \caption{\noindent Hadronic formation
time as a function of hadron mass for several quark p$_{T}$ and a
fixed fractional momentum z. The shaded areas show the upper and
lower limits for the lifetime of the partonic phase at RHIC and LHC,
respectively.} \label{qualitat}
\end{figure}

The qualitative dependence of leading hadron formation times on the
mass of the particle and the momentum and mass of the parent parton
are given in Figure~\ref{qualitat}. When compared to the lifetime of
the QGP at RHIC and the LHC we observe that for a wide range of
momenta, hadron formation times are well within the anticipated
lifetime of the plasma.

This is of particular interest for hadronic resonances, i.e. states
that might be formed and decay within the lifetime of the partonic
medium. Here, the hadron formed in the QGP will experience in-medium
interactions with the partonic system and decay off-shell if chiral
symmetry restoration reduces its mass. Consequently, intermediate momentum
resonances which are formed early and decay into particles that
escape the partonic medium with little hadronic reinteraction could be
used to test chiral symmetry restoration. According to UrQMD
calculations~\cite{urqmd}, hadronic reinteraction (rescattering and
regeneration) processes only affect the low momentum region ($p_{
T}$ = 0-2~GeV/c) of the resonance spectra. Therefore, for higher
momentum resonances, experimentally observable effects, such as
reduced production rates, modification of the branching ratios, mass
shifts and broadening of the widths in the QGP phase are expected
to be detectable. We take this as
an indication that a more detailed study of their production
and decay is justified to help guide experimental searches at RHIC
and the LHC.

\section{Formation time of heavy mass resonances in the partonic medium}

The study of particle formation in high energy heavy ion collisions requires
information about the distribution of the {\em observed} hadron momentum
fraction  relative to the parent quark or gluon, $P(z)$.  At  fixed momenta
$\{p_T\} = p_{T_1}\cdots p_{T_j}\cdots p_{T_n} $ and rapidities
$\{y\} = y_1 \cdots y_j \cdots y_n$, for a perturbative n-particle
production event,  including the simplest case of inclusive
hadrons, this probability is given by:
\begin{equation}
P(z_j) =  \frac{1}{ \frac{d\sigma^h(\{p_T\},\{y\})}{d\{y\} d \{p_T\} }  }
 \frac{d \sigma^h (\{p_T\},\{y\})}{d\{y\} d \{p_T \} \, dz_j}  \;.
\label{pz}
\end{equation}
The principal theoretical difficulty in obtaining $P(z)$ for resonances is
their unknown fragmentation functions, $D(z,Q^2)$.  Limited experimental
data on resonance production has hindered their inclusion in a global QCD
analysis~\cite{akk,soffer}. The approach that we here undertake is to use as a
guideline the known pion, kaon, proton, neutron and lambda decay
probabilities in approximating fragmentation into resonant states,
$D_{h^*}(z,Q^2) \propto D_h(z,Q^2) $. One notes that the overall
normalization of $D_{h^*}(z,Q^2)$ cancels in Eq.~(\ref{pz}) and it would
require decay probabilities of drastically different {\em shape}
to significantly alter $P(z)$ and, consequently, the hadron formation times for
$m_h > \Lambda_{QCD}$.

\begin{figure}[!t]
\includegraphics[width=2.9in]{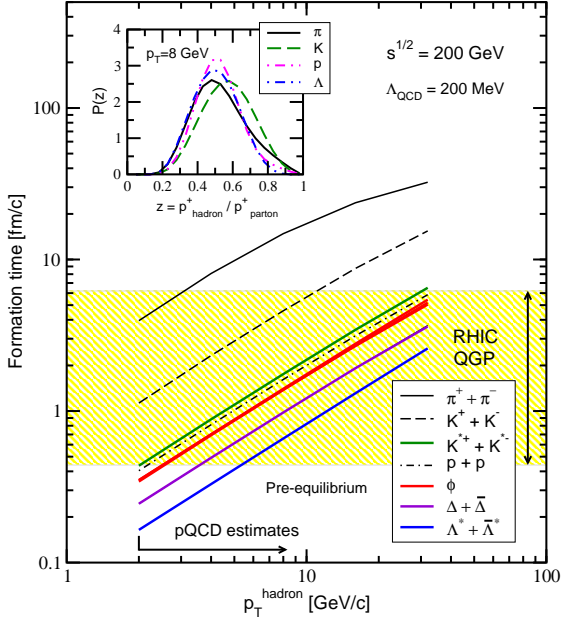}
\label{form1} \vspace*{.2cm} \caption{\noindent Transverse momentum
dependence of leading hadron formation times at RHIC energies,
$\sqrt{s}=200$~GeV. Results for both 'stable' particles and
resonances are presented. The shaded area represents the estimated
partonic lifetime at RHIC. Insert shows the meson and baryon
distributions $P(z)$ at a fixed momentum $p_T = 8$~GeV/c.}
\label{rhic}
\end{figure}

\begin{figure}[!b]
\includegraphics[width=2.9in]{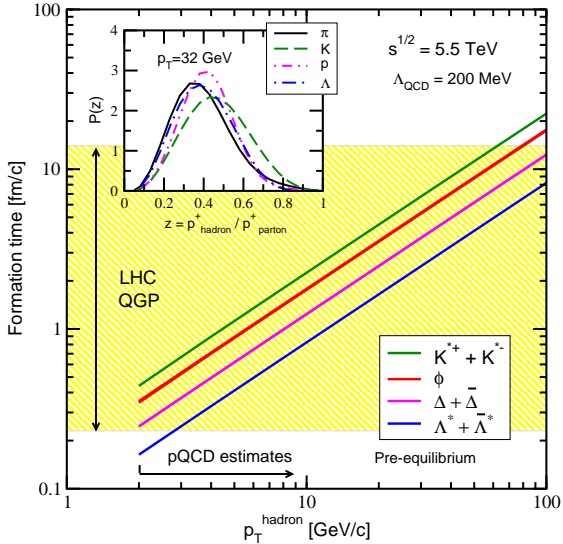}
\label{form2} \vspace*{.2cm} \caption{ \noindent Transverse momentum
dependence of heavy resonance formation times at LHC  energies,
$\sqrt{s}=5.5$~TeV. The shaded area represents the estimated
partonic lifetime at LHC. Insert shows the meson and baryon
distributions $P(z)$ at a fixed momentum  $p_T = 32$~GeV/c. }
\label{lhc}
\end{figure}

To lowest order, the perturbative QCD cross sections of hard,
$p_T >$~few~GeV/c,  inclusive and jet- or  hadron-triggered
($\Delta \varphi =  \varphi_2 -\varphi_1 \approx \pi$)
particle production are calculated as follows~\cite{Qiu:2004da,Vitev:2006bi}:
\begin{eqnarray}
\frac{ d\sigma^{h_1}_{NN} }{ dy_1  dp_{T_1}  }
& = &
K_{NLO}  \sum_{abcd} 2 \pi \,  p_{T_1}
\int_{x_{1,2}\leq 1}  dy_2 \int_{x_{1,2}\leq 1}  d z_1
   \nonumber \\
&& \hspace*{-2cm} \times \frac{1}{z_1^2} D_{h_1/c}(z_1)
   \frac{\phi_{a/N}({x}_a) \phi_{b/N}({x}_b) }{{x}_a{x}_b} \,
\frac{\alpha_s^2}{{S}^2 }  |\overline {M}_{ab\rightarrow cd}|^2   \;.
\label{single}
\end{eqnarray}

\begin{eqnarray}
\frac{ d\sigma^{h_1 h_2}_{NN} }{ dy_1  dy_2 d p_{T_1}  d p_{T_2} }
&=&
K_{NLO} \sum_{abcd} \,  2\pi  \int_{x_1 \leq 1, x_2 \leq 1,
z_2 \leq 1 }  d z_1
\nonumber \\
&& \hspace*{-1cm}
\times \frac{1}{z_1}  D_{h_1/c}(z_1)  D_{h_2/d}(z_2)
\frac{\phi_{a/N}({x}_a) \phi_{b/N}({x}_b) }{{x}_a{x}_b} \,
\nonumber \\
&& \hspace*{-1cm}
\times \frac{\alpha_s^2}{{S}^2 }
|\overline {M}_{ab\rightarrow cd}|^2   \;.
\label{double}
\end{eqnarray}
In  Eqs.~(\ref{single}) and (\ref{double}) vacuum and medium-induced
acoplanarity can also be incorporated~\cite{Vitev:2008vk}, but these have
a very small effect on the hadron formation in the kinematic region of
interest. It is important to note that for triggered back-to-back
correlations to lowest order $p_{T_1}/z_1 = p_{T_2} /z_2$~\cite{Rak:2004gk}.
For our choice of fragmentation functions, Ref.~\cite{akk}, the small
$z<0.05$ region has to be extrapolated.

Results for the mean formation time of hadronic resonances as  a  function
of their $p_T$,
\begin{equation}
\langle \tau_{\rm form} \rangle = \int_0^1 dz\, P(z)\,
\tau_{\rm form} (z; p_T,m_h,m_q,k_T=\Lambda_{QCD}) \; ,
\label{tauconv}
\end{equation}
compared to the estimated lifetime of the plasma at RHIC, are given
in Figure~\ref{rhic}. We note that at fixed  transverse momentum
$\langle  \tau_{\rm form}  \rangle $ is controlled primarily by the
particle mass. At higher $p_T$ there is a systematic bias toward
larger mean values of $z$ but these only affect the formation time
of the lightest particles, such as $\pi$ and $K$. We also studied
the validity of our assumption that we can get guidance for the
formation times of resonant particles from the known $P(z)$ momentum
fraction distributions for $\pi$, $K$, $p$, and $\Lambda$, shown in
the insert of Figure~\ref{rhic} for  $p_T = 8$~GeV/c at RHIC.
$\tau_{\rm form}$ for the  $\phi$ meson was evaluated with all
available momentum fraction distributions and the uncertainty,
presented in Figure~\ref{rhic} is not significant. In the
perturbative \ regime, $p_T > {\rm few}$~GeV/c light hadron production
takes place largely at times that exceed the QGP lifetime and the
observed nuclear modification is dominated by inelastic interactions
of the parent parton~\cite{Vitev:2005he}. Heavy particles, however,
tend to be produced inside the plasma and would undergo broadening
or dissociation~\cite{vitev}.  Depending on their $p_T$ range,
$K^*$, $\phi$, $\Delta$ and $\Lambda^*$  are ideally suited to prove
the early stages of the collision when the medium is hot and dense.
We finally note that in the region $\langle \tau_{\rm form} \rangle
\sim \tau_{\rm col.}  \ll \tau_0$ our formation time estimate
becomes less reliable since the separation of scales between hard
and soft physics, upon which this study is based, is violated.
Furthermore, only high momentum resonances ($p_{T}$ $>$ 2 GeV/c)
from the perturbative regime are considered, because low momentum
bulk resonances from a thermal QGP phase will be formed later and
are more likely, according to UrQMD simulation, to undergo large
rescattering in the hadronic medium.

Calculations of the formation times of the resonant states of
interest at the LHC are shown in Figure~\ref{lhc}. It is remarkable
that over a very large $p_T$ range (for $\Lambda^*$ almost to $p_T
\sim 100$~GeV/c) these hadrons can provide information for the
in-medium modification of particles in the background of hot dense
matter. Our conclusion is also facilitated by the longer lifetime of
the plasma created at the LHC. However, one has to take into account
that the plasma density and thus the resonance-parton interaction
probability will drop according to Bjorken expansion. Thus high
momentum resonances formed late in a rather dilute plasma will not
exhibit much of a medium modification. Furthermore the additional
decay time of the resonances has to be taken into account, therefore
a smaller momentum range will enrich the decays inside the dense
medium.

\section{Resonance interaction and decay in the partonic medium}

Medium modification of the resonance properties requires interaction
with the partonic medium. There are no detailed models of
parton-hadron interactions, but one can use an estimate of the
collisional wavefunction broadening for the propagating resonance,
characterized by $\Delta k_{T}^{2}$, $\Delta k_{T} = k_{\perp\, 1} - k_{\perp\, 2} $,
as has been done previously for propagating heavy mesons~\cite{vitev}. The amount
of broadening depends on the path length in the medium, the resonance formation time,
the resonance lifetime, and the time-dependent medium density. For a meson it
reads:
\begin{equation}
\Delta k_{T}^{2} = 4 \left\langle  \mu^2 \frac{L}{\lambda}  \right\rangle \xi
\propto \int_{\tau_{\rm form}}^{\tau_{\rm decay}} \alpha_s^2 \rho(z) dz  \;.
\label{broad}
\end{equation}
In Eq.~(\ref{broad})  $\mu$ sets the scale of typical momentum
transfers from the medium to a valence quark and $\xi$ simulates the
effect of the power law tails in Moliere multiple  scattering. One
notes that $\tau_{\rm decay} =( \tau_{\rm decay} )^* \gamma $
depends both on the in-medium modification of the resonance lifetime
and its momentum.

\begin{figure}[!t]
\includegraphics[width=2.7in]{zacor.eps} \\[3ex]
\includegraphics[width=3.2in]{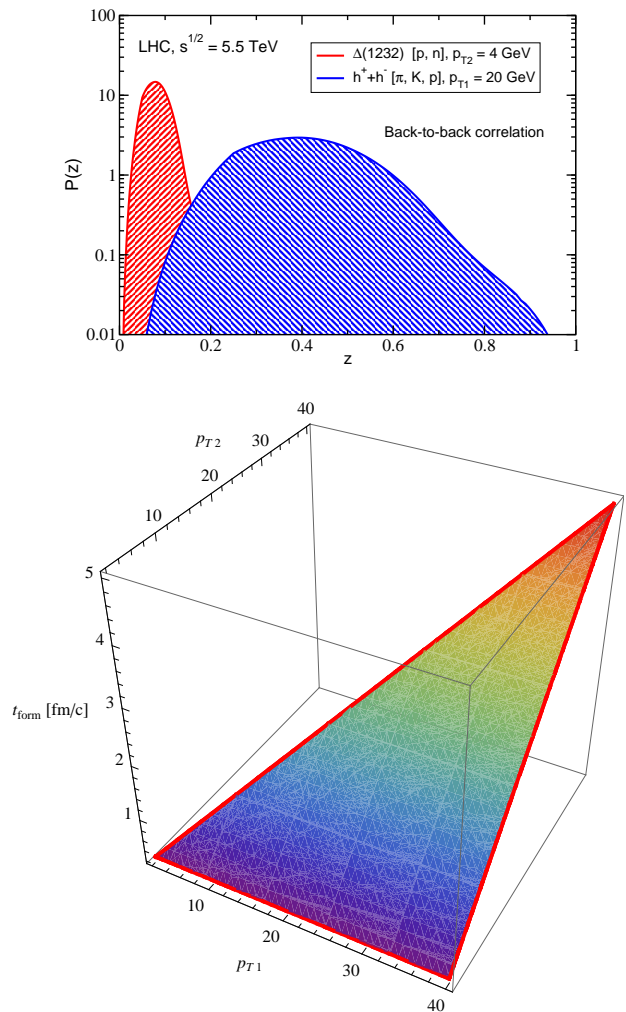}
\vspace*{.2cm}
\caption{\noindent Top panel: $P(z)$ distributions for the trigger $h^++h^-$
of $p_{T_1} = 20$~GeV/c and the associate away-side $\Delta (1232)$ of
$p_{T_2} = 4$~GeV/c in $\sqrt{s} = 5.5$~TeV  collisions at the  LHC.
Bottom panel: Formation time of $\Delta(1232)$ of momentum $p_{T_2}$,
triggered by charged hadron of momentum $p_{T_1}$, at the LHC. Results show
negligible dependence on the momentum of the leading trigger
hadron for massive correlated away-side resonances. }
\label{form3D}
\end{figure}

Given the small in-vacuum decay widths of heavy resonances,
$\Gamma(K^*) = 50.8$~MeV, $\Gamma(\phi) = 4.65$~MeV, $\Gamma(\Delta)
\approx 120$~MeV and $\Gamma(\Lambda^*) = 15.6$~MeV, in order to
observe any effect of partial chiral symmetry restoration  a
self-consistent reduction in their lifetime, related to the width
broadening, is critical. Otherwise, even if the partonic medium is
radially co-moving with velocity as large as $<\beta>_T \simeq
0.7c$~\cite{star-rad}, it is likely that high momentum resonances
with moderate in-vacuum lifetime (e.g. $K^*$, $\Lambda^*$) will have
their decay point boosted outside of the QGP. This sets the
constraint that the medium-modified lifetime of the resonance can
not exceed a few fm/c. Several models have been applied in the
past~\cite{Holt:2004tp,Filip:2001st}  to calculate the broadening of
the in-medium spectral functions  and, although these calculations
are based on interactions in a dense hadronic medium, parton-hadron
duality arguments can hopefully  allow to generalize the results to
a quark-gluon plasma. It has been argued~\cite{Holt:2004tp}, on the
example of the $\phi$ meson, that at $T=250$~MeV the shortening of its
in-vacuum lifetime can be as large as a factor of 10.

Furthermore, care should be taken in ensuring that the resonance propagates through
an extended volume of hot and dense matter. Triggering on a strongly interacting
particle biases the partonic hard scattering toward the surface of the fireball and
guarantees that the away-side resonance must traverse a
medium of larger spatial extent.  The  higher  the trigger $p_T$, the more effective
this technique is expected to be. Ideally, one needs two handles to control
$\tau_{\rm form}$ and the trigger bias. Naively, one may expect that energetic hadron
triggers that lead to sizable partonic $p^+$ will entail resonance formation outside
of the QGP, see Eq.~(\ref{tfrag}). However, the selection of an associated $p_T$
correlates the near-side and away-side momentum fractions, $p_{T_2}/p_{T_1}
\approx z_2/z_1 $. This relation is exact  to lowest order  and is  illustrated
for back-to-back particles of significantly different momenta, on the example
of their probability distributions $P(z_1)$ and $P(z_2)$,  in the top panel of
Figure~\ref{form3D}. The bottom panel of Figure~\ref{form3D} shows that for massive
hadrons, $m_h \gg \Lambda_{QCD}$, the resonance momentum  determines its formation
time independent of the trigger $p_T$.

The 10-fold broadening of the $\phi$ meson decay width in medium at
T=250~MeV, which was modeled in \cite{Holt:2004tp}, implies $\Delta
\Gamma_{th} \approx 50$~MeV. Using this thermal contribution to the
decay width as a rough guidance, one can obtain estimates for the
reduced lifetime of hadronic resonances. Subsequent accounting for
the Lorentz time dilation and the time dependence of the QGP
temperature at RHIC and LHC suggests that experimental studies
should focus on hadronic resonances below 5 GeV/c (RHIC) or 10 GeV/c
(LHC), respectively, assuming that the lifetime of the partonic
medium has to at least equal the total resonance propagation time,
including formation and decay. Due to the convolution of many
dynamical effects, a more quantitative estimate on how many of the
produced resonances will decay in the medium is near impossible at
present. For example, there is a time distribution of formation and
decay, $\propto [ 1 - \exp(-t/\tau_{\rm form, \; decay}) ]$,
suggestive that there will always be a small fraction of resonances,
affected by the medium. But as has been the case in all previous
studies, short lifetime resonances are more likely to exhibit a
larger effect. Therefore studies of the K$^{*}$ (c$\tau$ = 3.88 fm)
and $\Delta$ (c$\tau$ = 1.64 fm) will yield stronger signatures than
the $\Lambda^{*}$ (c$\tau$ = 12.6 fm) and the $\phi$ (c$\tau$ = 46.5
fm).

Finally, we address the question of dissociation of lightly bound
resonances in the partonic medium, see e.g Eq.~(\ref{broad}).
Theoretically, this may lead to very strong resonance suppression,
but experimental evidence at RHIC~\cite{1,2} has shown that although
the resonant over non-resonant particle ratios decrease in going
from proton-proton to central Au+Au collisions, the survival rate of
direct resonance production is large. It is difficult to quantify
the rate because regeneration of resonances in the hadronic phase
has also been experimentally verified \cite{1}. A semi-quantitative
analysis of direct $\phi$-production vs. $\phi$-production from $K^+
+ K^-$ coalescence has shown that for longer lived resonances the
hadronic contribution is small \cite{3}, but the relative strength
of the hadronically generated  resonance contribution changes with
lifetime and hadronic interaction probabilities \cite{4}. Recent
measurements by PHENIX in the di-lepton sector \cite{5} might be a
first indicator that resonances are indeed medium modified in the
hot partonic phase at RHIC. These results need to be evaluated in
accordance with other RHIC meaurements by STAR, though, which have
shown that cold nuclear matter effects already modify the in-vacuum
resonance properties, albeit on a small scale \cite{2,6}.

\section{An experimental method to search for chiral
symmetry restoration}

In order to experimentally verify the impact of the partonic medium
on resonance properties we need to reconstruct intermediate momentum
resonances which are produced early. This might be possible through
the selection of resonances in the away-side jet of a triggered
di-jet or $\gamma$-jet event. The trigger can be based on a
reconstructed high momentum (leading) particle or the full near-side
jet energy reconstructed in a calorimeter. Both methods lead to the
identification of a near-side jet which is less affected by the
medium due to the surface bias in the trigger. The
($\Delta\phi=\pi$) correlations with the away-side associated
particles will determine the medium modified jet. Full jet
reconstruction on the near side is preferred, because it determines
not only the exact $z$-distribution of the away-side jet fragments
but also the jet axis. However, back-to-back jet simulations based
on the reconstruction of leading trigger particles already show that
the overall $z$-distribution of the leading particle on the
near-side constrains the $z$-distribution of the associated
particles on the away-side, as shown if Figure~\ref{form3D}.

\begin{figure}[hbtl]
\includegraphics[width=2.9in]{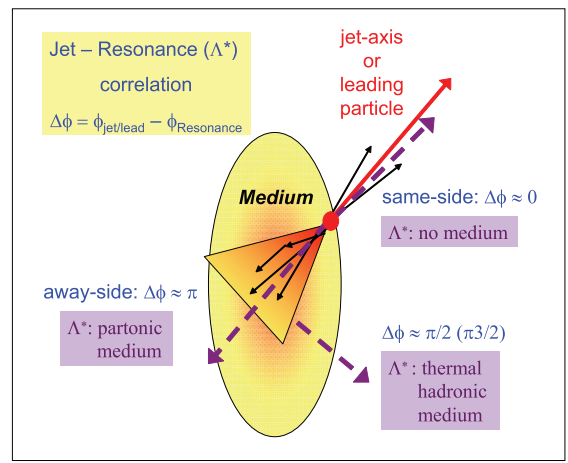}
 \caption{Sketch of jet fragmentation into
resonances ($\Lambda$* is taken as an example for all hadronic
resonances) in the medium created in a heavy-ion collision.
Same-side correlations of resonances are not affected by the medium,
whereas the away-side high p$_{\rm T}$ resonance might be affected
by the early-time (chirally restored) medium. Thermal resonances,
which are affected by the late-time hadronic medium are at $\pi$/2
with respect to the trigger particle.}
 \label{jetresosketch}
\end{figure}

In order to study the chiral transition we propose a $\Delta\phi$
quadrant analysis(see Figure~\ref{jetresosketch}).

The medium modification of resonance properties should be strongest
in the away-side ($\Delta \phi = \pi$) quadrant using reconstructed
resonances with momentum of p$_{T}$ $>$ 2 GeV/c which are correlated
to the same-side jet axis or leading particle. They are likely to be
produced early and thus will interact with the partonic medium, but
their decay products will also escape the medium sufficiently fast
to not exhibit much interaction in the late hadronic phase.

\begin{figure}[h!]
\includegraphics[width=3.5in]{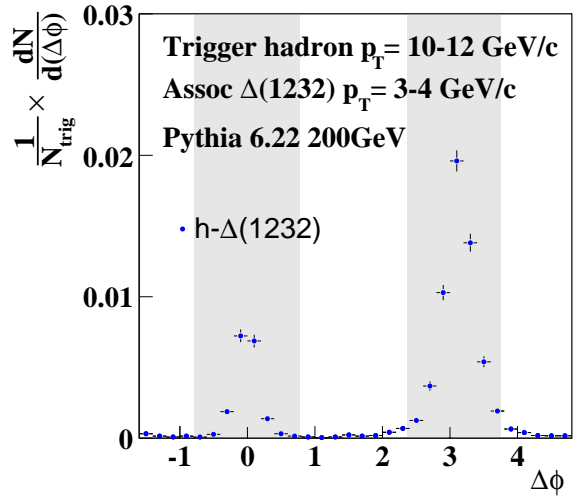}
 \caption{PYTHIA simulation of a hadron-$\Delta$ correlation in proton-proton
collisions for typical hadron trigger and associated particle
kinematics.}
 \label{jetphidist}
\end{figure}

The low momentum resonances produced at an angle of $\Delta \phi =
1/2 \pi$ or $3/2 \pi$ with respect to the jet axis or leading
particle are identified as thermal resonances produced late from the
bulk matter of the collision. They do not populate the spectrum in
proton-proton collisions, as shown in Figure~\ref{jetphidist}, but
they will be a dominant source in A+A collisions. In the proposed
analysis they will be used as a reference, because they
predominantly interact in the late hadronic medium. Therefore, their
masses and widths are expected to not be altered much relative to
the vacuum, unless phase space effects from late regeneration change
the shape of the invariant mass signal.

The high p$_{\rm T}$ same-side jet resonances are also expected to
exhibit the vacuum width and mass because the initial quark is
expected to fragment outside of the medium (surface bias effect).
Therefore a differential measurement of high momentum resonances as
a function of the angle to the jet axis might have a built-in
reference system and may not require comparison to measurements in
p+p collisions, which are very statistics limited for the rare
hadronic resonance channels.

Figure~\ref{jetphidist} shows a simulation of the $\Delta\phi$
distribution for correlations between a trigger hadron and
associated $\Delta$ resonances based on PYTHIA. The four quadrants
proposed in Figure~\ref{jetresosketch} are highlighted in the plot.
Reconstructing the invariant mass in these specific $\Delta\phi$
ranges will yield information on mass shifts, width broadening and
branching ratios. Quantitative studies of these properties as a
function of resonance momentum, emission angle, jet energy, and jet
tag, will directly address the question of chiral symmetry
restoration in collisions of heavy nuclei. Early studies of this
method have been reported recently based on a statistics limited
data set at RHIC \cite{markert2}. \\

\section{Summary and Conclusions}

The enhanced jet cross sections at the LHC  will  enable  the  heavy
ion experimentalists to carry out an in-depth triggerable intermediate-
and high-momentum resonance correlation program, which will scan the
relevant kinematic domain to look for chiral symmetry restoration
effects in hadronic resonances through the method described in this
paper. We recognize the inherent uncertainties associated with the
estimates of the formation and decay of hadrons in a highly relativistic,
possibly strongly-coupled, quark-gluon plasma at high density and
temperature, the possibility that resonances can be dissociated in the
medium, and the finite probability that even at intermediate and high
momentum hadronic rescattering of their decay products might mimic
QGP effects. Nevertheless, this is the most complete study to date which
uses theoretical input to guide the experimental searches for the
chiral transition in QCD in conjunction with the deconfinement phase
transition, one of the original drivers of the heavy ion programs at
SPS, RHIC and the LHC. Based on our results, we  anticipate that this
method can yield a definitive measurement of the modification (or
lack thereof) of the properties of hadronic resonances in the QGP.
The fact that our proposed technique has a built-in reference by
simply analyzing off-axis low momentum resonances in the same event
makes it even more attractive experimentally.

\begin{acknowledgments}
We thank B.W. Zhang, H. Huang and C. Greiner for useful discussions.
This work is supported by the U.S. Department of Energy  Office of
Science under contracts No. DE-AC52-06NA25396, DE-FG02-94ER40845,
and DE-FG02-92FR40713.
\end{acknowledgments}


\begin{thebibliography}{80}

\bibitem{bial} e.g. A. Airapetian et al. (HERMES coll.), {\it Phys.
Lett.} {\bf B557}, 37 (2003).

\bibitem{accardi} A. Accardi, {\it Phys. Lett.} {\bf B649}, 384
(2007).

\bibitem{cassing} T. Falter, {\it Phys. Rev.} {\bf C70}, 054609
(2004).

\bibitem{vitev}
  A.~Adil and I.~Vitev,
  {\it Phys. Lett.} {\bf B649}, 139 (2007)


\bibitem{heavy1} S.S. Adler et al. [PHENIX coll.],
{\it Phys. Rev. Lett.} {\bf 94}, 082301 (2005).

\bibitem{heavy2} B.I. Abelev et al. [STAR coll.],
{\it Phys. Rev. Lett.} {\bf 98}, 192301 (2007).


\bibitem{Holt:2004tp}
  L.~Holt and K.~Haglin,
  {\it J. Phys.} {\bf G31}, S245 (2005).


\bibitem{Filip:2001st}
  P.~Filip and E.~E.~Kolomeitsev,
  {\it Phys. Rev.} {\bf C64}, 054905 (2001).

\bibitem{Cassing:1997jz}
  W.~Cassing, E.~L.~Bratkovskaya, R.~Rapp and J.~Wambach,
  Phys.\ Rev.\  C {\bf 57}, 916 (1998).

\bibitem{baryons}
  L.~Ya.~Glozman,
  {\it Phys. Lett.} {\bf B475}, 329 (2000).


\bibitem{Bjorken:1976mk}
  J.~D.~Bjorken,
  {\it Lect. Notes Phys.}  {\bf 56}, 93 (1976).


\bibitem{lund} B.Andersson et al., {\it Phys. Rep.} {\bf 97}, 31
(1983).

\bibitem{Kopeliovich:1984ur}
  B.~Z.~Kopeliovich and F.~Niedermayer,
  Sov.\ J.\ Nucl.\ Phys.\  {\bf 42}, 504 (1985)
  [Yad.\ Fiz.\  {\bf 42}, 797 (1985)].



\bibitem{gyulassy} A. Bialas and M.Gyulassy, {\it Nucl. Phys.} {\bf B291}, 793
(1987).


\bibitem{falter} K. Gallmeister, T. Falter, {\it Phys. Lett.} {\bf B630},
40 (2005).



\bibitem{akk} S.Albino, B.A. Kniehl, G. Karmer, {\it Nucl. Phys.} {\bf B734}, 50
(2006).
\bibitem{soffer} C. Bourrely, J. Soffer, {\it Phys. Rev.} {\bf D68}, 014003
(2003).



\bibitem{Vitev:2007ve}
  I.~Vitev,
  {\it Phys. Rev.} {\bf C75}, 064906 (2007).


\bibitem{Vitev:2005he}
  I.~Vitev,
  {\it Phys. Lett.} {\bf B639}, 38 (2006).



\bibitem{Adler:2003cb} S.~S.~Adler et al.  [PHENIX Coll.],{\it Phys. Rev. Lett.} {\bf
91} 072301 (2003).

\bibitem{Sjostrand:2006za} T.~Sjostrand, S.~Mrenna and P.~Skands, {\it JHEP} {\bf 0605}, 026 (2006).


\bibitem{markert1} B.I. Abelev et al., {\it Phys. Rev. Lett.} {\bf 97} 132301 (2006).

\bibitem{urqmd} M. Bleicher et al.,  {\it Phys. Lett.} {\bf  B530}, 81
(2002).


\bibitem{Qiu:2004da}
  J.~W.~Qiu and I.~Vitev,
  {\it Phys. Lett.}{\bf B632}, 507 (2006).


\bibitem{Vitev:2006bi}
  I.~Vitev, J.~T.~Goldman, M.~B.~Johnson and J.~W.~Qiu,
  {\it Phys. Rev.}{\bf D74}, 054010 (2006).


\bibitem{Vitev:2008vk}
  I.~Vitev and B.~W.~Zhang,
  arXiv:0804.3805 [hep-ph].



\bibitem{Rak:2004gk}
  J.~Rak,
  {\it J. Phys.}{\bf G30}, S1309 (2004).


\bibitem{star-rad}
J. Adams et al., [STAR coll.], {\it Phys. Rev. Lett.} {\bf 92},
112301 (2004).

\bibitem{1}
B.I. Abelev et al. [STAR collab.], {\it Phys. Rev. Lett.} {\bf 97},
132301 (2006).

\bibitem{2}
B. I. Abelev et al. [STAR collab.], arXiv:0801.0450 [nucl-ex].

\bibitem{3}
J. Adams et al. [STAR collab.], {\it Phys. Lett.} {\bf B612}, 181 (2005).

\bibitem{4}
C. Markert, {\it J. Phys.} {\bf G31}, S169 (2005).

\bibitem{5}
S. Afanasiev et al. [PHENIX coll.], arXiv:0706.3034 [nucl-ex].

\bibitem{6} J. Adams et al. [STAR coll.], {\it Phys. Rev. Lett.}
{\bf 92}, 092301 (2004).

\bibitem{markert2} C. Markert, arXiv:0706.0724 [nucl-ex].

\end{thebibliography}
\end{document}